\documentclass[letter]{aa} 

\usepackage{graphicx}
\usepackage{txfonts}
\usepackage[utf8]{inputenc} 
\usepackage{color,soul}
\usepackage{amsmath}

\begin{document}

   \title{Accuracy of meteor positioning from space- and ground-based observations}


   \author{Hongru Chen
          \inst{1,2}
          \and
          Nicolas Rambaux\inst{1}
          \and
          Jérémie Vaubaillon \inst{1}
          }

   \institute{IMCCE, Observatoire de Paris, CNRS, Université PSL, Sorbonne Université, Université de Lille, 77 av. Denfert-Rochereau, 75014 Paris, France \\
	\email{hongru.chen@obspm.fr; nicolas.rambaux@obspm.fr; jeremie.vaubaillon@obspm.fr}
	\and
	Planetary Environment and Asteroid Resource Laboratory, Origin Space Co. Ltd., 5 Shihua Rd, 518048 Shenzhen, China\\
	\email{hongru.chen@origin.space}
	%
}


 
  \abstract
   {
}
   {The knowledge of the orbits and origins derived from meteors is important for the study of meteoroids and of the early solar system. With an increase in nano-satellite projects dedicated to Earth observations or directly to meteor observations (e.g., the Meteorix CubeSat), we investigate the stereoscopic measurement of meteor positions using a pair of cameras, one deployed in space and one on the ground, and aim to understand the accuracy and the main driving factors. This study will reveal the requirements for system setups and the geometry favorable for meteor triangulation.}
   { 
   	 This Letter presents the principle of the stereoscopic measurement from space and the ground, and an error analysis. Specifically, the impacts of the resolutions of the cameras, the attitude and orbit determination accuracy of the satellite, and the geometry formed by the moving target and observers are investigated.

}
   {To reach a desirable positioning accuracy of 1 km it is necessary to equip the satellite with high-accuracy sensors (e.g., star tracker and GPS receiver) to perform fine attitude and orbit determination. The best accuracy can occur when the the target is at an elevation of $30^\circ$ with respect to the ground station. }
   {}
	

   \maketitle
%

\section{Introduction}

Meteoroids are remnants of the early solar system. When a meteoroid enters Earth's atmosphere, it generates a luminous trace known as a meteor. Determining meteor trajectories allows us to determine the origin of the meteoroid \citep[e.g., parent body and age,][]{1981Icar...45..545D,1986MNRAS.219...47O,2006mspc.book.....J,2017A&A...598A..15S,2019AA...622A..84G,2019msme.book.....R}. In addition, knowledge of meteoroid flux density can constrain environment models, such as the IMEM2, which helps us to plan safe space activities in the entire solar system \citep{1993JGR....9817029D,Lemcke1997,MCBRIDE19971513,2019AA...628A.109S}. Knowledge of the flux density near Earth is especially important for planetary defense. Many ground-based tracking networks (e.g., FRIPON, CAMS, DFN, EDMOND, and AMOS) have been developed with the purpose of determining the flux density \citep{2014acm..conf..110C,2018EPSC...12.1105C,2015PSS..118...38R,2015PSS..118..102T,2016SASS...35..165D,2016Icar..266..384J}. Moreover, space-borne observation of meteors has recently been investigated. The space-proof camera SPOSH was developed for meteor detection \citep{2011PSS...59....1O}. Two projects have been launched with the purpose of observing meteors from space, namely the SCUBE CubeSat and the METEOR experiment on the International Space Station (ISS) \citep{2014LPI....45.1846I,2019LPICo2157.6493A}. The CubeSat project Meteorix is planned for detecting meteors and re-entering space debris~\citep{rambaux:insu-01851524,rambaux:hal-02198139}. 

As a meteor is a transient event, the reconstruction of its orbit generally requires observations from distributed stations. General methods for meteor triangulation have been well documented \citep[e.g.,][]{ 1967codp.book.....W}. Scenarios and algorithms that employ multiple images taken from space only or from the ground only have been intensively studied \citep{1959PhT....12k..54O,1981pmp..book.....B,1987BAICz..38..222C,1990BAICz..41..391B,GURAL2012,2017PSS..143...89H,2017Icar..294...43E,2020MNRAS.491.2688V,2019Icar..321..388S, Chen2020}. However, a combination of space- and ground-based observations has not yet been explored. With the establishment of country- to continental-scale meteor networks 
and the emergence of space-borne meteor observations
, the chance to capture meteors from both the ground and space is increasing with time. 
Moreover, any Earth observation satellite can participate in meteor observations in conjunction with ground stations; the number of Earth observation satellite is increasing thanks to the establishment of new satellite constellations (e.g., the Starlink project).
In particular, the future Meteorix can work with any station of ground networks (e.g., FRIPON, CAMS, and DFN) to enable and enhance meteor orbit determination.  
The long baseline between the satellite and the ground station seems favorable in terms of stereoscopic measurements. Nevertheless, the positioning accuracy is subject to the geometry formed by the moving target and observers; the setup of the systems, such as the attitude and orbit determination accuracy of the satellite; and image resolution. 

This Letter presents an error analysis against the above-mentioned factors, and shows the requirements for system setups and optimal geometry. After briefly recalling the space- and ground-based meteor observations (section \ref{sec:obsbasics}), we present the principle of stereoscopic measurement and methods of deriving position errors for the scenario of interest (section \ref{sec:stereo}), the results showing the influences of different error sources as well as the necessary system setup (section \ref{sec:results1}), and a discussion on the impact of the observation geometry (section \ref{sec:results2}).

\section{Basic observation settings}\label{sec:obsbasics}

Concerning the space-borne observation, we refer to the design of the Meteorix project. The CubeSat is equipped with a camera and a detection chain that features near real-time detection at low power consumption~\citep{10.1145/2870650.2870652}. The camera is placed at the end of the 3U CubeSat and pointed toward the nadir. 
The science orbit is set to a sun-synchronous orbit, which is the most common orbit that maximizes launch opportunities. It is assumed that the orbit is nearly circular with an altitude around 500 km. The period is 94.7 min. 
The local time of the ascending node is set to 10:30. Consequently, the night time lasts around 32 min per orbit. It has been estimated that this setup allows a detection of around 100 meteors in one year ~\citep{rambaux:hal-02198139}. 
The selected image sensor has 2048$\times$2048 pixels, and the camera has a field of view of 40$^\circ$.  
For the ground-based observation, we adopt the specification of a current FRIPON station: field of view of 180$^\circ$ (like many other all-sky networks, such as CAMS and AMOS), resolution of roughly 1000$\times$1000 pixels, and axis of the camera pointed toward the zenith. 


The usual observable altitude of meteors lies between 80 and 120 km \citep{1998SSRv...84..327C}. The speed of a meteor with respect to the ground is in the range from 11 to 72 km/s. Considering a frame rate of 10 fps, to capture the motion of the slowest meteors the position accuracy should be better than 1.1 km. 

\section{Space and ground stereoscopic measurement}\label{sec:stereo}

\subsection{Principle}
\begin{figure}
	\centering\includegraphics[trim=0 0 20cm 0, clip,width=3.5in]{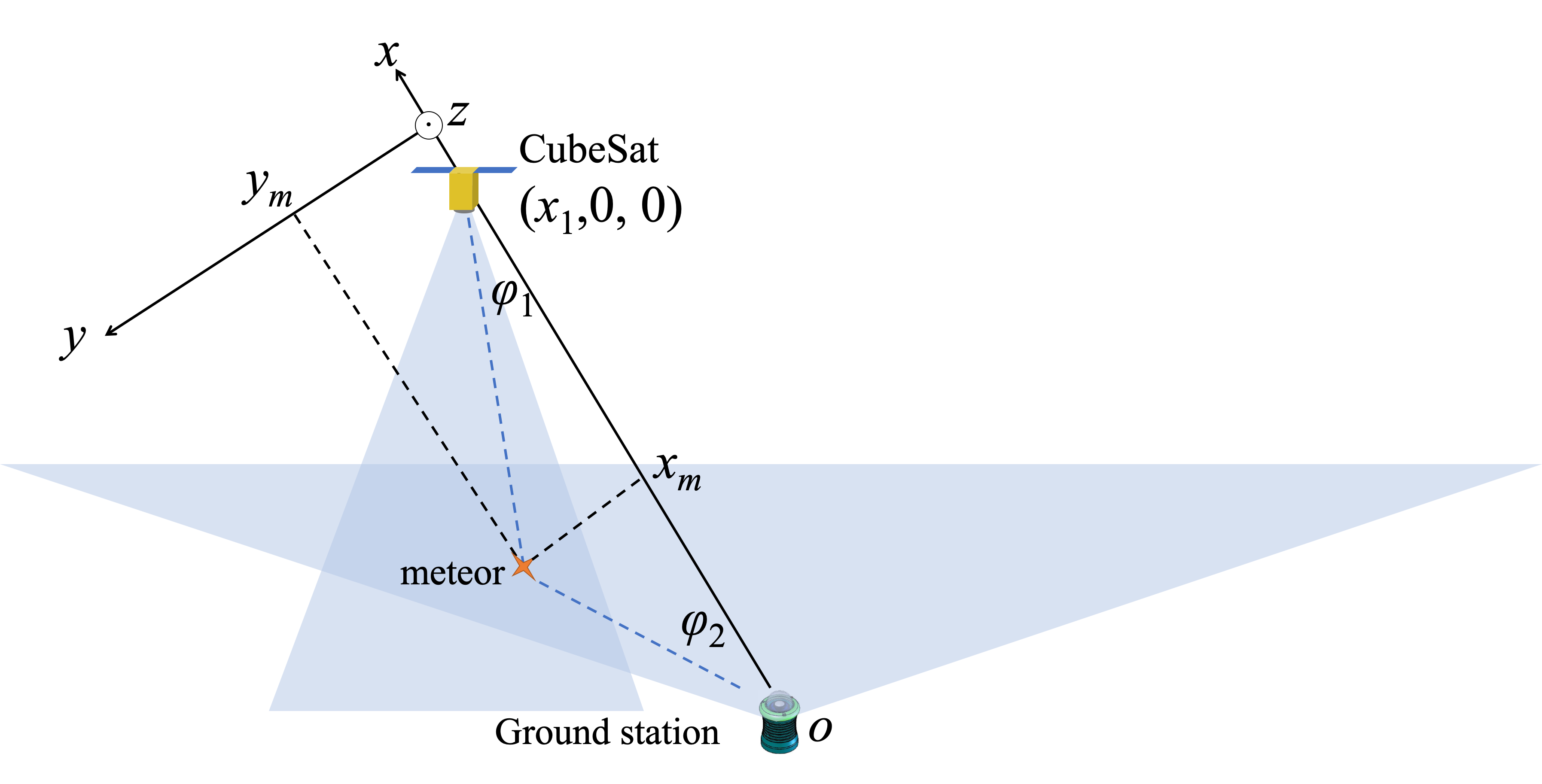}
	\caption{Stereoscopic measurement by a CubeSat and a ground station, and the defined coordinate system.}
	\label{fig:3dMesure}
\end{figure}

With only one camera the direction from the camera to the target can be measured, while the distance cannot be. With simultaneous observations by stations distributed in space (e.g., Meteroix CubeSat) and on the ground (e.g., FRIPON), the distance between the target and the two stations as well as the three-dimensional coordinates can be determined, as depicted in Fig.~\ref{fig:3dMesure}. Without loss of generality, 
the coordinate system is chosen such that the origin, $O$, coincides with the location of the ground station, the $x$-axis points to the CubeSat, the $y$-axis is perpendicular to the $x$-axis and aligned with the direction to the meteor, and the $z$-axis completes the right-handed coordinate system. Let $\left[x_1, y_1, z_1\right]^{\text{T}}$ denote the coordinate of the CubeSat. The true coordinate will be $\left[x_1, 0, 0\right]^{\text{T}}$. Let $\left[x_m, y_m, z_m\right]^{\text{T}}$ denote the coordinate of the targeted meteor, and $\bf{i}$, $\bf{j}$, and $\bf{k}$ denote the unit vector along the $x$-, $y$-, and $z$-axis, respectively. The position vector of the meteor is
\begin{equation}\label{Eq:pos}
{\bf{r}}={\bf i}x_m+{\bf j}y_m+{\bf k}z_m
.\end{equation}
Let $\varphi_1$ and $\varphi_2$ represent angles between the directions from the observers to the meteor and the baseline between the two observers, as depicted in Fig.~\ref{fig:3dMesure}. The direction from an observer to the target can be derived from the two-dimensional coordinates on the image. The coordinates of the meteor can be computed from
\begin{equation}
\begin{gathered}\label{eq:coord}
x_m = \frac{{{x_1}}}{{C + 1}},\hfill \\
y_m = \frac{{{x_1}\tan {\varphi _2}}}{{C + 1}},\hfill \\
z_m =0,\hfill\\
\end{gathered}
\end{equation} 
where
\begin{equation}
C=\frac{\tan{\varphi_2}}{\tan{\varphi_1}}
.\end{equation}

\subsection{Measurement errors}
The attitude and orbit determination errors of the CubeSat and image resolutions will translate into errors of coordinates of the meteor. Differentiating Eq.(\ref{Eq:pos}) yields
\begin{equation}\label{Eq:diff_pos}
\delta{\bf{r}}={\bf i}\delta x_m+x_m\delta{\bf i}+{\bf j}\delta y_m+y_m\delta{\bf j}+{\bf k}\delta z_m+z_m\delta{\bf k}
.\end{equation}
Considering only in-plane (i.e., the $x$-$y$ plane) errors, orbit determination error $\delta y_1$ along ${\bf j}$ results in a change of the reference frame, which is expressed as
\begin{equation}
\begin{gathered}
\delta{\bf i}=\frac{\bf j}{B}\delta y_1,  \hfill\\
\delta{\bf j}=-\frac{\bf i}{B}\delta y_1, \hfill\\
\end{gathered}
\end{equation}
where $B=x_1$, representing the length of the baseline between the two cameras. For out-of-plane errors the last term of Eq.(\ref{Eq:diff_pos}) is zero as $z_m$ is zero. Let $\delta{\bf{r}}$ also be expressed as
\begin{equation}\label{Eq:diff_pos2}
\delta{\bf{r}}={\bf i}\delta r_i+{\bf j}\delta r_j+{\bf k}\delta r_k
,\end{equation}
where
\begin{equation}
\begin{gathered}\label{eq:rearrange}
\delta r_i = \delta x_m - \frac{y_m}{B}\delta y_1,\hfill \\
\delta r_j = \delta y_m + \frac{x_m}{B}\delta y_1,\hfill \\
\delta r_k = \delta z_m. \hfill
\end{gathered}
\end{equation}

Differentiating $x_m$ and $y_m$ in Eq.~(\ref{eq:coord}) with respect to the variables yields
\begin{equation}
\begin{gathered}\label{eq:diff}
\delta x_m = \frac{1}{{C + 1}}\delta {x_1} + \frac{{B\tan {\varphi _2}}}{{{{(C + 1)}^2}{{\sin }^2}{\varphi _1}}}\delta {\varphi _1} - \frac{{B{{\sec }^2}{\varphi _2}}}{{{{(C + 1)}^2}\tan {\varphi _1}}}\delta {\varphi _2}, \hfill \\
\delta y_m = \frac{{\tan {\varphi _2}}}{{C + 1}}\delta {x_1} + \frac{{B{{\tan }^2}{\varphi _2}}}{{{{(C + 1)}^2}{{\sin }^2}{\varphi _1}}}\delta {\varphi _1} + \frac{{B{{\sec }^2}{\varphi _2}}}{{{{(C + 1)}^2}}}\delta {\varphi _2}.\hfill \\
\end{gathered}
\end{equation} 
The direction error $\delta \varphi_1$ is a result of in-plane image resolution, $\delta\varphi^{\parallel}_{res1}$; the in-plane attitude determination error, $\delta\varphi^{\parallel}_{att}$; and the position error of the CubeSat along the $y$-axis, which is expressed as
\begin{equation}
\delta\varphi_1=\delta\varphi^{\parallel}_{res1}+\delta\varphi^{\parallel}_{att}+\frac{1}{B}\delta y_1 
.\end{equation}
Similarly, $\delta \varphi_2$ is extended to
\begin{equation}
\delta\varphi_2=\delta\varphi^{\parallel}_{res2}-\frac{1}{B}\delta y_1 
,\end{equation}
where $\delta\varphi^{\parallel}_{res2}$ represents the in-plane direction error due to the image resolution of the ground station. 

It is tricky to determine the offset $\delta z_m$ along ${\bf k}$ when the measured directions from the cameras to the target do not intersect in three-dimensional space due to errors. Let $\delta z_{I}$ and $\delta z_{II}$ denote the out-of-plane errors attributed to the CubeSat and the ground station, respectively. The $z$ coordinate error $\delta z_m$ can be approximated by
\begin{equation}\label{eq:diffz1}
\delta z_m = \frac{\delta z_{I}+\delta z_{II}}{2}
.\end{equation}
Here $\delta z_{I}$ is a result of the out-of-plane image resolution, $\delta\varphi^{\perp}_{res1}$; attitude determination error, $\delta\varphi^{\perp}_{att}$; and  orbit determination error, $\delta z_1$, which is expressed as
\begin{equation}
\delta z_{I} = \frac{{BC}}{{(C + 1)\cos {\varphi _{1}}}}(\delta{\varphi^{\perp} _{res1}} +\delta{\varphi^{\perp} _{att}})+{\delta z_1}
,\end{equation}
and $\delta z_{II}$, due to the out-of-plane image resolution of the ground station, $\delta\varphi^{\perp}_{res2}$, is expressed as
\begin{equation}\label{eq:zii}
\delta z_{II} = \frac{B}{{(C + 1)\cos {\varphi _{2}}}}\delta {\varphi^{\perp}_{res2}}
.\end{equation}

It can be reasonably assumed that in-plane and out-of-plane direction errors vary independently.  Uncertainties of errors have the relationship,
\begin{equation}
\begin{gathered}
\sigma\varphi^{\parallel}_{res1}=\sigma\varphi^{\perp}_{res1}=\left(\frac{{{\theta _1}}}{{px_1}}\right),\hfill \\
\sigma\varphi^{\parallel}_{res2}=\sigma\varphi^{\perp}_{res2}=\left(\frac{{{\theta _2}}}{{px_2}}\right),\hfill \\
\sigma{\varphi^{\parallel}_{att}}=\sigma{\varphi^{\perp} _{att}}, \hfill \\
\end{gathered}
\end{equation}
where $\sigma$ indicates the 1$\sigma$ uncertainty of the variable following it, $\theta_1$ and $\theta _2$ respectively denote the fields of views, and $px_1$ and $px _2$ respectively denote the pixels (in a line) of the CubeSat and the ground station. It is also assumed that distribution of orbit determination error in space is homogeneous:
\begin{equation}
\sigma x_1=\sigma y_1 = \sigma z_1
.\end{equation}

The position uncertainty $\sigma r$ is defined as the root sum of variances of the three position components, which is expressed as  
\begin{equation}\label{eq:RSSerror}
\sigma r = \sqrt{{\sigma ^2}{r_i}+{\sigma ^2}{r_j}+{\sigma ^2}{r_k}}
,\end{equation}
where $\sigma^2$ indicates the variance of the variable following it.
According to Eq.(\ref{eq:rearrange}-\ref{eq:zii}) and given that the error sources are uncorrelated, ${\sigma ^2}{r_i}$, ${\sigma ^2}{r_j}$, and ${\sigma ^2}{r_k}$ can be computed from
\begin{equation}\label{eq:error_ri}
\begin{aligned}
{\sigma ^2}{r_i} ={} & {\left( {\frac{1}{{C + 1}}} \right)^2}{\sigma ^2}{x_1} + {\left( {\frac{{B\tan {\varphi _2}}}{{{{(C + 1)}^2}{{\sin }^2}{\varphi _1}}}} \right)^2}({\sigma ^2}\varphi _{res1}^{||} + {\sigma ^2}\varphi _{att}^{||})\\
 &+ {\left( {\frac{{B{{\sec }^2}{\varphi _2}}}{{{{(C + 1)}^2}\tan {\varphi _1}}}} \right)^2}{\sigma ^2}\varphi _{res2}^{||}\\
 &+ {\left( {\frac{{\tan {\varphi _2}}}{{{{(C + 1)}^2}{{\sin }^2}{\varphi _1}}} + \frac{{{{\sec }^2}{\varphi _2}}}{{{{(C + 1)}^2}\tan {\varphi _1}}} - \frac{{\tan {\varphi _2}}}{{(C + 1)}}} \right)^2}{\sigma ^2}{y_1}
,\end{aligned}
\end{equation}
\begin{equation}\label{eq:error_rj}
\begin{aligned}
{\sigma ^2}{r_j} = {} & {\left( {\frac{{\tan {\varphi _2}}}{{C + 1}}} \right)^2}{\sigma ^2}{x_1} + {\left( {\frac{{B{{\tan }^2}{\varphi _2}}}{{{{(C + 1)}^2}{{\sin }^2}{\varphi _1}}}} \right)^2}({\sigma ^2}\varphi _{res1}^{||} + {\sigma ^2}\varphi _{att}^{||})\\
 &+ {\left( {\frac{{B{{\sec }^2}{\varphi _2}}}{{{{(C + 1)}^2}}}} \right)^2}{\sigma ^2}\varphi _{res2}^{||}\\
 &+ {\left( {\frac{{{{\tan }^2}{\varphi _2}}}{{{{(C + 1)}^2}{{\sin }^2}{\varphi _1}}} - \frac{{{{\sec }^2}{\varphi _2}}}{{{{(C + 1)}^2}}} + \frac{1}{{(C + 1)}}} \right)^2}{\sigma ^2}{y_1}
,\end{aligned}
\end{equation}
\begin{equation}\label{eq:error_rk}
\begin{aligned}
{\sigma ^2}{r_k} = {} &{\left( {\frac{{BC}}{{2(C + 1)\cos {\varphi _1}}}} \right)^2}({\sigma ^2}\varphi _{res1}^ \bot  + {\sigma ^2}\varphi _{att}^ \bot )\\
 &+ {\left( {\frac{B}{{2(C + 1)\cos {\varphi _2}}}} \right)^2}{\sigma ^2}\varphi _{res2}^ \bot + \frac{1}{4}{\sigma ^2}{z_1}
.\end{aligned}
\end{equation}

\subsection{Reduction of geometries}
\begin{figure}
	\centering\includegraphics[trim=2.5cm 0 2.2cm 0, clip, width=3.5in]{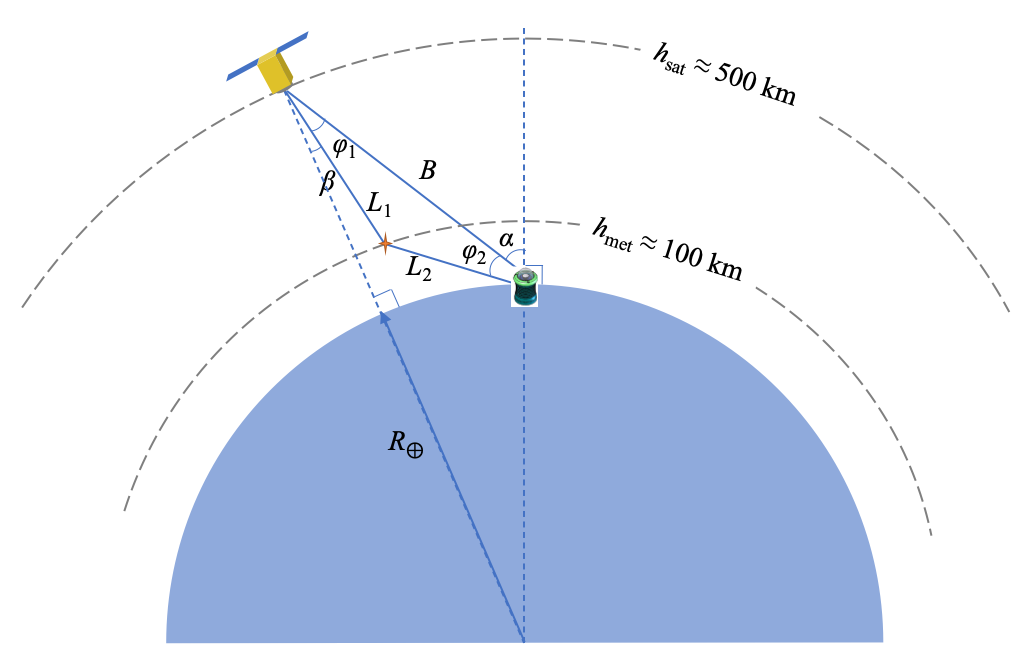}
	\caption{Simplified geometry for meteor observation.}
	\label{fig:Geometry}
\end{figure}

Figure~\ref{fig:Geometry} depicts the geometry of space- and ground-based observations. Altitudes of the CubeSat, $h_{sat}$, and meteors, $h_{met}$, vary within small ranges (around 500 km and between 80 and 120 km, respectively). In the following analyses, $h_{sat}$ is set to 500 km and $h_{met}$ 100 km to represent a general situation. The observation geometry can then be defined by the directions from the ground station to the CubeSat and the meteor. 
For convenience, the direction of the CubeSat with respect to the zenith of the ground station, $\alpha$, and the direction of the observed meteor from the CubeSat with respect to the camera axis, $\beta$, are chosen to generalize the observing geometry of interest. As $\alpha$ and $\beta$ are limited by the fields of view of the two observers, possible observation geometries can thus be represented by an $\alpha$-$\beta$ map. Taking into account the round surface of the Earth (where the radius of the Earth $R_ \oplus = 6371$ km is used), $B$, $\varphi_1$, and $\varphi_2$ entering in Eq.~(\ref{eq:error_ri}-\ref{eq:error_rk}) are computed as follows. First,
\begin{equation}
B =  - {R_ \oplus }\cos \alpha  + \sqrt {{{({R_ \oplus } + {h_{sat}})}^2} - {R_ \oplus }^2{{\sin }^2}\alpha }
.\end{equation}
The triangle formed by the observers and target does not necessarily lie on the vertical plane. The angle $\varphi_1$ is actually within an interval expressed as
\begin{equation*}
{\varphi _1} \in \left[ {\arcsin \left( {\frac{{{R_ \oplus }\sin (\pi  - \alpha )}}{{{R_ \oplus } + {h_{sat}}}}} \right) - \beta ,\arcsin \left( {\frac{{{R_ \oplus }\sin (\pi  - \alpha )}}{{{R_ \oplus } + {h_{sat}}}}} \right) + \beta } \right]
.\end{equation*}
As $\beta$ is to be varied in the range of the field of view (e.g., -20$^\circ$ to 20$^\circ$ in this work), which can cover situations at two ends of the intervals, the problem can be reasonably simplified by fixing $\varphi_1$ for a given $\beta$ to 
\begin{equation}
{\varphi _1} = \arcsin \left( {\frac{{{R_ \oplus }\sin (\pi  - \alpha )}}{{{R_ \oplus } + {h_{sat}}}}} \right) - \beta
.\end{equation}
Then
\begin{equation}
{\varphi _2} = \arcsin \left(\frac{{{L_1}}}{{{L_2}}}\sin {\varphi _1}\right)
,\end{equation}
where the distances from the meteor to the CubeSat, $L_1$, and the ground station, $L_2$, are computed from
\begin{equation}
\begin{gathered}
{L_1} = ({R_ \oplus } + {h_{sat}})\cos \beta  - \sqrt {{{({R_ \oplus } + {h_{met}})}^2} - {{({R_ \oplus } + {h_{sat}})}^2}{{\sin }^2}\beta } , \hfill \\
{L_2} = \sqrt {{L_1}^2 + {B^2} - 2{L_1}B\cos {\varphi _1}}. \hfill
\end{gathered}
\end{equation}

\section{Results}\label{sec:results1}
In order to provide a guideline for system setups, we look at the influences of the orbit determination error (e.g., $\sigma x_1$), attitude determination error $\sigma {\varphi _{att}}$, image error of the CubeSat $\theta_1/px_1$, and image error of the ground station $\theta_2/px_2$. 
The baseline setup assumes $\sigma x_1= 10$ km, corresponding to coarse orbit determination based on TLE data only \citep{Smith2019}, $\sigma {\varphi _{att}}=4^\circ$, corresponding to a coarse attitude determination based on magnetometer and sun sensors only \citep{KULeuven2018}, $\theta_1=40^\circ$, $px_1=2048$, $\theta_2=180^\circ$, and $px_2=1000$, corresponding to the baseline design of Meteorix and FRIPON. Then $\alpha$ ranges from 0 to 90$^\circ$, and $\beta$ from -20$^\circ$ to 20$^\circ$.

\begin{figure*}
	\centering
	\includegraphics[trim=3cm 0 2cm 0, clip, width=17cm]{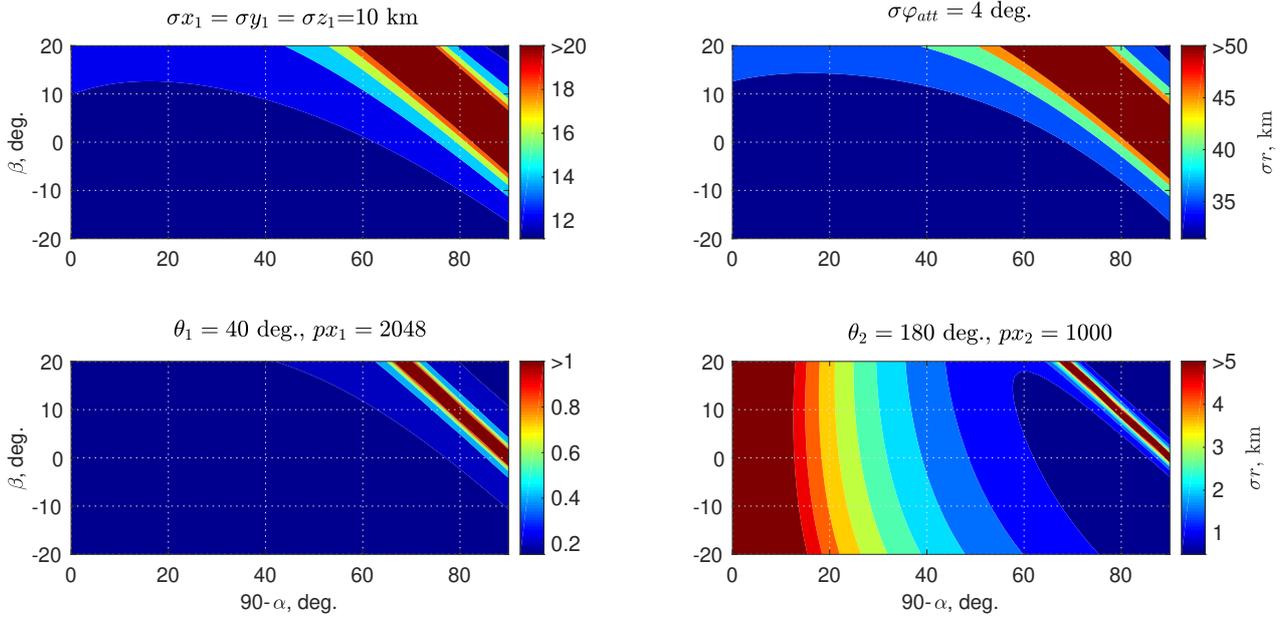}
	\caption{Positioning uncertainty (color levels) contributed by baseline orbit and attitude determination errors (upper panels) of the CubeSat, and baseline image resolution of the CubeSat and the ground station (lower panels).}
	\label{fig:ErrorComponents}
\end{figure*}

\begin{figure*}
	\centering
	\includegraphics[width=15cm]{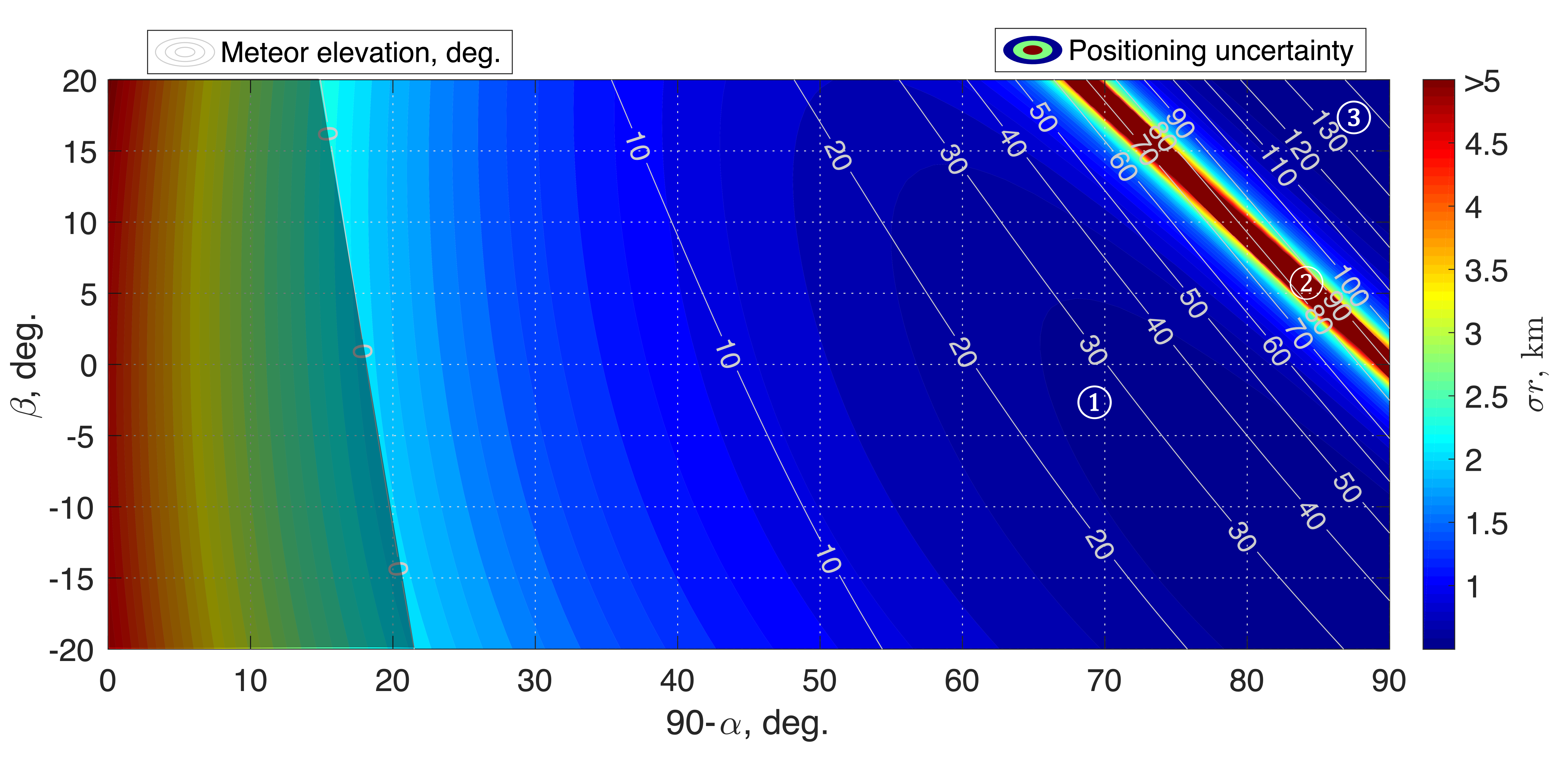}
	\caption{Positioning uncertainty (color levels) for the recommended system setup, and elevation (gray lines) of the meteor with respect to the ground station.}
	\label{fig:GeometryDependent}
\end{figure*}

The position accuracy $\sigma r$ computed from Eq.~(\ref{eq:RSSerror}-\ref{eq:error_rk}) is presented in Fig.~\ref{fig:ErrorComponents} as a function of $90^\circ-\alpha$ (i.e., elevation of the CubeSat) and $\beta$. The four panels correspond to the cases considering orbit determination error only (upper left), attitude determination error only (upper right), image error of the CubeSat only (lower left), and image error of the ground station only (lower right). 
The lower left panel shows that the CubeSat image resolution is good enough to not cause $\sigma r>$ 0.5 km, except in the singular-geometry region  (see next section for discussion). The upper right panel shows that coarse attitude estimation results in a $\sigma r$ around 40 km for most observation geometries. Therefore, the CubeSat should perform fine attitude estimation where a star tracker is necessary. In this case, $\sigma {\varphi _{att}}=0.04^\circ$ is expected \citep{KULeuven2018}, and the contribution of $\sigma {\varphi _{att}}$ to $\sigma r$ decreases to 0.4 km. Even with this improvement the resulting $\sigma r$ can still go beyond 15 km, due to the orbit determination error (see the upper left panel). It is recommended to perform fine orbit determination as well. A GPS receiver can enable a fine orbit determination at an accuracy of 0.1 km \citep{Smith2019}, which will reduce the error contribution to 0.15 km for most geometries. The resulting $\sigma r$ can go up to 2 km for some situations, given the effect of the image resolution of the ground station. To widely achieve a desired $\sigma r$ of 1.1 km, it is recommended to increase the image resolution of the ground station, for instance to 2048$\times$2048 pixels. With an upgraded resolution, the minimum $\sigma r$ will go down to 1 km for most geometries. The position uncertainty $\sigma r$ following the recommended setups is shown in Fig.~\ref{fig:GeometryDependent}. The pattern is similar to the lower left panel of Fig.~\ref{fig:ErrorComponents}, which means the contribution of $\sigma \varphi_2$ due to the image resolution of ground station is dominant.

\section{Discussion}\label{sec:results2}
Figure~\ref{fig:GeometryDependent} also displays the elevation of the meteor with respect to the ground station, which is derived from $\alpha$ and $\beta$. The shaded area where the meteor is below the horizon is not a valid area for observation, and thus is excluded from our discussion. In the figure, an elevation $>90^\circ$ (which is not a common usage) indicates that the meteor and the CubeSat are on different sides of the zenith of the ground station.

Figure~\ref{fig:GeometryDependent} shows that $\sigma r$ strongly depends on the elevation of the meteor. Following the recommended setup, $\sigma r<2$ km is widely achieved except for the elevation interval $\left[65^\circ,90^\circ\right]$ (labeled case 2 in the figure).
The corresponding geometry is referred to as the singular geometry. When the meteor is at an elevation of around $30^\circ$ (labeled case 1) and also captured by the CubeSat camera, the geometry is optimal with $\sigma r$ ranging from 0.5 to 0.8 km, which depends on the observing directions from the CubeSat.  Another optimal region appears in the elevation interval $\left[110^\circ,140^\circ\right]$ (labeled case 3), which corresponds to the commonly used elevation $\left[40^\circ, 70^\circ\right]$ with the CubeSat observing from the other side of the zenith. 

\begin{figure}
	\centering\includegraphics[trim=0.8cm 0 0.5cm 0, clip, width=3.5in]{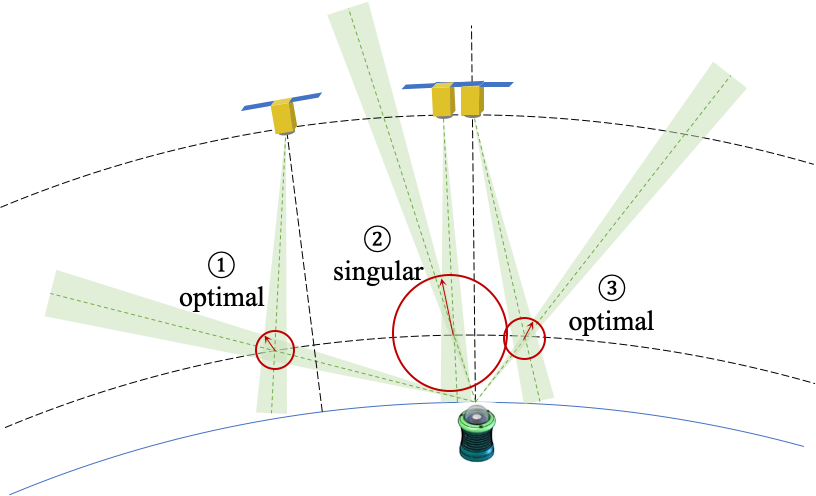}
	\caption{Intersecting error zones depending on the geometry.}
	\label{fig:GeometryEffect}
\end{figure}

Figure~\ref{fig:GeometryEffect} schematically visualizes the singular and optimal geometries revealed in Fig.~\ref{fig:GeometryDependent}. With only one camera, the error (green band) resulting from the angular size of a pixel is infinite along the observing direction (green dotted line), which should be constrained by the observation from another direction. If two directions are aligned, the error is still infinite. Because of the limited field of view of a CubeSat, it will be roughly on top of the target when capturing the target. When the meteor is near the zenith of the ground station (case 2 in Fig.~\ref{fig:GeometryEffect}) and the CubeSat observes it from the top, the error along the observing direction is not effectively constrained. Therefore, the resulting radius of the error zone (i.e., the intersecting green zone) is large. This explains the singularity associated with the meteor elevation interval $\left[65^\circ,90^\circ\right]$. When the elevation of the meteor is low ($30^\circ$ in case 1 and $40^\circ$ in case 3), the included angle between the two observing directions is close to 90$^\circ$,  leading to effective triangulation and small errors.
Even though the error {across} the observing direction is proportional to the square of the distance, and thus large for low-elevation observations, the resulting range of errors, which is mainly contributed by the error {along} the observing direction, is still minimized for low elevations. 

\section{Conclusions}
      To achieve a desirable positioning accuracy of 1 km, the CubeSat should bring high-accuracy sensors (e.g., star tracker and GPS receiver) to perform fine attitude (i.e., accuracy $\sim0.05^\circ$) and orbit (i.e., accuracy $\sim$ 100 m) determination. 
      In addition, the image resolution of FRIPON is recommended to upgrade to 2048$\times$2048 pixels. The performance of meteor triangulation is best (i.e., with smallest position error) when the meteor elevation is around $30^\circ$. With the recommended system setup, as long as a meteor is captured by both the CubeSat and the ground station, and at an elevation $<65^\circ$ with respect to the ground station, the position and the orbit of the meteor can be determined.


\begin{acknowledgements}
	We acknowledge the ESEP (Exploration Spatiale des Environnements Planétaires) postdoctoral fellowship, DIM ACAV+ R\'egion \^Ile-de-France, Janus CNES, and IDEX Sorbonne Universit\'es for funding this research.
\end{acknowledgements}

%
%

   \bibliographystyle{aa} 
   \bibliography{references.bib} 

\end{document}